\documentstyle[12pt]{article}

 2
\def\BI{{\rm 1\!l}}

\def\Eq#1{{\begin{equation} #1 \end{equation}}}

\begin{document}
\begin{flushright}
UAHEP 961\\
January 1996\\
\end{flushright}

\centerline{ \LARGE   Bosonic Description of Spinning Strings }
\centerline{ \LARGE    in $2+1$   Dimensions}
\vskip 2cm

\centerline{ {\sc  B. Harms and A. Stern }}

\vskip 1cm

\centerline{  Dept. of Physics and Astronomy, Univ. of Alabama,
Tuscaloosa, Al 35487, U.S.A.}

\vskip 1cm

\vskip 2cm

\vspace*{5mm}

\normalsize
\centerline{\bf ABSTRACT}

\vskip 2cm
\vspace*{5mm}

We write down a general action principle for spinning strings in
$2+1$ dimensional space-time {\it
without introducing Grassmann variables}.  The action is written solely
in terms of
coordinates taking values in the $2+1$ Poincar\'e group, and it
has the usual string symmetries, i.e. it is  invariant under {\it a})
 diffeomorphisms of the world sheet and {\it b}) Poincar\'e
  transformations.   The system can
be generalized to an arbitrary number of space-time dimensions,
and also to spinning membranes and p-branes.

\newpage
\scrollmode
\setcounter{footnote}{1}
\baselineskip=24pt

It is well known that the classical spin of  relativistic particles
can be described by using either classical or pseudoclasssical
variables.\cite{rbok}   Here we show that the same holds for spinning
strings.  Since the psuedoclassical description of spinning strings
is well known,   we obviously are implying  that
there exist descriptions of spinning strings solely in
terms of classical variables.  For reasons of simplicity,
we shall  examine strings in $2+1$ space-time only.  (The string action
has a particularly elegant form in $2+1$ dimensions due to the existence
of a nondegenerate scalar product on the Poincar\'e algebra
$\underline{ISO(2,1)}$.\cite{wit})  Our system
contains the general description of (spinless) strings due to
 Balachandran, Lizzi
and Sparano\cite{bal} as a special case.\footnote{\scriptsize
    Spinning strings were
also considered in \cite{bal} using a Wess-Zumino term.  Here we
shall show that there are more possibilities for including spin.}
Furthermore, it can
be generalized to an arbitrary number of space-time dimensions,
and also to spinning membranes and p-branes.  We shall discuss such
generalizations in a later article.

Analogous to the bosonic formulation of a spinning particle
in $2+1$ dimensions (cf. \cite{ss}), we write the spinning string action
 on the   $2+1$
dimensional Poincar\'e group manifold $ISO(2,1)$.  That is, the
string variables are maps from
the two dimensional world sheet to $ISO(2,1)$.
We shall express the action for strings in
$2+1$ dimensional Minkowski space in terms of these variables and their
derivatives.  It will be seen to be invariant under {\it a})
diffeomorphisms of the world sheet and
{\it b}) global Poincar\'e transformations.  The strings can be
classified in terms of $ISO(2,1)$ orbits, and we shall find that certain
orbits correspond to strings  with a nonvanishing spin current.
Lastly, we shall show how to
embed such strings in curved space-time,
 the resulting action being invariant under {\it a}),
and now {\it b}$'$) local Poincar\'e transformations.

We denote the  string variables by $g$.
 $g$ can be decomposed into an $SO(2,1)$ matrix $\Lambda=\{{\Lambda^i}_j\;,\;
i,j=0,1,2\}$ and an $SO(2,1)$  vector $x=\{x^i\;,\;i=0,1,2\}$.
The latter will
denote the Minkowski coordinates of the string.   Under the left action
of the Poincar\'e group, $g=(\Lambda, x)$
 transforms according to the semidirect product
rule: \Eq{g\rightarrow h \circ g = (\theta,y)\circ (\Lambda,x)
  =(\theta\Lambda, \theta x +y)\;.\label{lapg}}

We let $t_i$ and $u_i,\;i=0,1,2$ denote a basis for the Lie-algebra
 $\underline{ISO(2,1)}$.  For their commutation relations we can take
\Eq{[t_i,t_j] = \epsilon_{ijk} t^k\;,\quad
{}[t_i,u_j] = \epsilon_{ijk} u^k\;, \quad
{}[u_i,u_j] = 0\;,}
where we raise and lower indices using the Minkowski metric
$[\eta_{ij}] =$diag$(-1,1,1)$, and we define the totally antisymmetric
tensor $\epsilon_{ijk}$ such that $\epsilon^{012}=1$.

A left invariant
Maurer-Cartan form can be expanded in this basis as follows:
\Eq{g^{-1}dg={1\over2}\epsilon^{ijk}(\Lambda^{-1}d\Lambda)_{ij} t_k +
(\Lambda^{-1}dx)^iu_i\;.}  It is easy to check that $g^{-1}dg$ is
unchanged under the left action of the Poincar\'e group (\ref{lapg}).
Under the right action of the Poincar\'e group,
\Eq{g\rightarrow g \circ h^{-1}= (\Lambda,x)\circ (\theta^{-1},-\theta^{-1}
y)   =(\Lambda\theta^{-1},x- \Lambda\theta^{-1}y)\;,}  and
consequently the Maurer-Cartan form transforms as
\Eq{g^{-1}dg\rightarrow {}^h [g^{-1}dg]
={1\over2}\epsilon^{ijk}(\Lambda^{-1}d\Lambda)_{ij} \;{}^h[t_k] +
(\Lambda^{-1}dx)^i\; {}^h[u_i]\;,}
where \Eq{
{}^h[t_i]= {\theta^j}_i t_j +\epsilon^{jk\ell}\theta_{ki}y_j
 u_\ell \;,\quad {}^h[u_i]= {\theta^j}_i u_j \;.}  Here
 ${}^h[t_i]$ and ${}^h[u_i]$ denote basis vectors which are
 transformed under the adjoint action by $h\in ISO(2,1)$.
They are given explicitly for $h=(\theta,y)$.
These equations can be utilized to
define the adjoint action $v\rightarrow {}^h [v]$
 by $h\in ISO(2,1)$ on any vector $v =\alpha^it_i
+\beta^i u_i$  in   $\underline{ISO(2,1)}$.

Two scalar products exist on $\underline{ISO(2,1)}$
 which are invariant under the above defined adjoint action.
   We denote them by $<\; ,\; >$ and ( , ).  The former satisfies
\Eq{<t_i,u_j>=\eta_{ij} \;,\quad <t_i,t_j>= <u_i,u_j>=0\;,} and is
nondegenerate.  The latter satisfies
\Eq{(t_i,t_j)=\eta_{ij} \;,\quad (u_i,t_j)= (u_i,u_j)=0\;,} and is
degenerate.   Thus for any two vectors $v$ and $v'$ in
 $\underline{ISO(2,1)}$ we have that $<{}^h[v],{}^h[v']>$ $=<v,v'>$ and
$({}^h[v],{}^h[v'])$  $=(v,v')$.

 The nondegenerate  scalar product $<\; ,\; >$  was utilized  previously
in writing down the action for a relativistic spinning particle.\cite{ss}
The expression for the action is linear in  the Maurer-Cartan form
and it is therefore  invariant under (left)
 Poincar\'e transformations, as well as diffeomorphisms
of the particle world line.  The action
is just \Eq{S_{particle} =  \int <K, g^{-1}dg>\;,  }
where here $g$ is a function of the world line and
$K$ is a constant vector in   $\underline{ISO(2,1)}$.
The  direction of $K$ in  $\underline{ISO(2,1)}$
determines whether the particle is massive, massless or tachyonic, and
whether it is  spinning or spinless. Actually, for this purpose,
 it is sufficient to specify the $ISO(2,1)$ orbit on which $K$ lies.
This is because both $K=K_0$ and $K={}^h[K_0]$, $h\in ISO(2,1)$,
lead to the same classical equations of motion.  This follows from
the invariance property of the scalar product $<\;,\;>$,
\Eq{ <K, g^{-1}dg> =  <{}^h[K],{}^h[ g^{-1}dg]>
 =  <{}^h[K], g'^{-1}dg' >  \;,  }
 where      $g'=   g \circ h^{-1}  \;. $
Thus the action is invariant under  $K\rightarrow {}^h[K]$ and the
change of coordinates $g \rightarrow g'$.  Now to specify the $ISO(2,1)$
orbit we can use the two invariants
$ <K,K>\;{\rm and} \; (K,K)\;.$   Spin is associated with the former
invariant, and we shall find an analogous result for strings as well.

For the case $K=m t_0 - \kappa u_0$, we end up with the known\cite{ss}
 bosonic description of a massive spinning
particle.  That choice corresponds to both invariants being nonvanishing:
$ <K,K>=m\kappa$ and $(K,K)=-m^2\;.$
For this case the action can be expressed according to
\Eq{<K, g^{-1}dg>=m {\Lambda^i}_0 dx_i+
\kappa   (\Lambda^{-1}d\Lambda)_{12}\;.}

The equations of motion for the particle are easy to obtain.
Transforming $g$ according to: $g\rightarrow
(1+\epsilon) \circ g$, where $\epsilon$ is an infinitesimal element of
$\underline{ISO(2,1)}$, induces the following variation of the
Maurer-Cartan form:
\Eq{\delta (g^{-1} dg) = {}^{g^{-1}}[d\epsilon] \;.\label{dmc}}   Then
\Eq{\delta S_{particle} =\int <K, {}^{g^{-1}}[d\epsilon]>
=\int <{}^g[K], d\epsilon> =-\int <d({}^g[K]), \epsilon> \;,}
where we have used the invariance property of the scalar product.
The equations of motion thus state that ${}^g[K]=p^i t_i - j^i u_i
$ are constants of the motion.  Upon once again choosing
$K=m t_0 - \kappa u_0$, we get the following expressions for these
constants: \Eq{
 p^i =  m{\Lambda^i}_0 \;,\quad
 j^i = m \epsilon^{ijk} x_j {\Lambda_k}_0 + \kappa {\Lambda^i}_0 \;.}
   The former can be
identified with the momenta of the particle, while the latter can be
identified with the angular momenta, the first term being the orbital
angular momenta and the second being the spin. The spin is thus
 proportional to $ \kappa$ which is nonvanishing when  $ <K,K>$ is.

We now apply an analogous procedure to the
description of spinning strings.   Once again we shall express
the action  in terms of the Maurer-Cartan form
and consequently it will be invariant under (left)
Poincar\'e transformations.
The action should be quadratic in $g^{-1} dg$
in order for it to also be invariant under diffeomorphisms
of the world sheet.   We
then take the tensor product of two such
Maurer-Cartan forms and write
\Eq{S_{string} =  \int <{\cal K}, g^{-1}dg \otimes g^{-1}dg >\;.
\label{stac} }
Now ${\cal K}$ is a constant tensor with values in $\underline{ISO(2,1)}
\otimes  \underline{ISO(2,1)}$.  Analogous to what happens for the
 particle action,
it is sufficient to specify the $ISO(2,1)$ orbit on which ${\cal K}$ lies.
This is because both ${\cal K}={\cal K}_0$ and
${\cal K}={}^h[{\cal K}_0]$, $h\in ISO(2,1)$,
lead to the same classical equations of motion, which is once again due to
the invariance property of the scalar product $<\;,\;>$,
\Eq{ <{\cal K}, g^{-1}dg\otimes g^{-1}dg> =
 <{}^h[{\cal K}],{}^h[ g^{-1}dg] \otimes {}^h[ g^{-1}dg]>
 =  <{}^h[{\cal K}], g'^{-1}dg'\otimes g'^{-1}dg' >
  \;,} where $ g'=   g \circ h^{-1} \;.   $
Thus the action is invariant under  ${\cal K}\rightarrow {}^h[{\cal K}]$
 and the
change of coordinates $g \rightarrow g'$.  Now to specify the $ISO(2,1)$
orbit we can use the two invariants
$ <{\cal K},{\cal K}>\;{\rm and} \; ({\cal K},{\cal K})\;.$

 The string action (\ref{stac}) has already been studied\cite{bal} for the case
 ${\cal K}=\epsilon^{ijk} n_k\; t_i \otimes t_j$ .
For that choice \Eq{<{\cal K}, g^{-1}dg \otimes g^{-1}dg >=
\epsilon_{ijk}(\Lambda n)^i\; dx^j\; dx^k\;.\label{inte}}
  It remains to specify  the constant vector $n_i$.  For
$n_{i}$ space-like, light-like or time-like
one recovers the Nambu string, the null string\cite{bal},\cite{nul}
 or the tachyonic
string, respectively.  Thus to get the Nambu string we may write,
$n_i={1\over{4\pi\alpha'}}\;\delta_{i2}$.  To see how this works
we may extremize the action with repect to the  variations of $\Lambda$:
$\delta\Lambda_{ij} =\epsilon_{ik\ell}
 {\Lambda^k}_j \zeta^\ell$, $\zeta^\ell$ being infinitesimal.  This
leads to the equations of motion
\Eq{\epsilon^{ijk} \Lambda_{j2} V_k =0\;,
\quad V_k =\epsilon_{ijk}dx^i\; dx^j\;.}  Then $V_i$ is parallel to
  $\Lambda_{i2}$.  Upon fixing the normalization, we get
$\Lambda_{i2}={V_i}/{\sqrt{V_j V^j}}\;.$  Substituting this
result back into the integrand (\ref{inte})
 yields the usual form for the Nambu action,
\Eq{S_{Nambu} ={1\over {4\pi\alpha'}}\; \int \sqrt {V_j V^j}
= {1\over {2\pi\alpha'}}\; \int  d^2\sigma\; \sqrt{ (\partial_0 \vec{x})^2
 (\partial_1 \vec{x})^2 -(\partial_0 \vec{x}\cdot\partial_1 \vec{x})^2}
\;,} where $\sigma=(\sigma_0,\sigma_1)$ parametrize the world sheet
and $\partial_0 ={\partial\over{\partial\sigma_0}}$ and
$\partial_1  ={\partial\over{\partial\sigma_1}}$.

We now return to the general form for the string action (\ref{stac}).
The equations of motion are obtained in identical fashion as was done
for the particle.  Transforming $g$ according to: $g\rightarrow
(1+\epsilon) \circ g$, where $\epsilon$ is an infinitesimal element of
$\underline{ISO(2,1)}$, once again induces the variation (\ref{dmc})
 of the  Maurer-Cartan form.  Then
\Eq{\delta S_{string} =2\int <{\cal K}, {}^{g^{-1}}[d\epsilon]\otimes
g^{-1}dg> =2\int <{}^g[{\cal K}], d\epsilon\otimes dgg^{-1}>\;,}
where we have used the invariance property
of the scalar product and the identity
\Eq{     {}^{g^{-1}}[dg g^{-1}]= g^{-1} dg \;.\label{lvsr}}
Upon integrating by parts we arrive at the equations of motion
\Eq{ d\; <{}^g[{\cal K}], T_A\otimes dgg^{-1}>=0\;,}  where the $T_A$'s
denote the  generators of $ ISO(2,1) $.
These equations state that there are six conserved currents.  For
$T_A=u_i$, we have  \Eq{ \partial_\alpha P_i^\alpha=0\;,\quad\;
P_i^\alpha=\epsilon^{\alpha\beta} <{}^g[{\cal K}], u_i\otimes
\partial_\beta gg^{-1}>\;,\label{cop}} which we identify with the momentum
current conservation.
  (Here $\alpha,\beta,...$ denote world sheet indices.)
  For $T_A=t_i$, we have  \Eq{ \partial_\alpha J_i^\alpha=0\;,\quad\;
J_i^\alpha=\epsilon^{\alpha\beta} <{}^g[{\cal K}], t_i\otimes
\partial_\beta gg^{-1}>\;,\label{coj}} which we identify with the angular moment
current conservation.

We now examine the conserved currents $P_i^\alpha$ and $J_i^\alpha$
for several cases.

\underbar{Case 1}:  ${\cal K}={\cal K}_1
=\epsilon^{ijk} n_k\; t_i \otimes t_j$.
This is the case we considered earlier which contains the Nambu string,
as well as the null and tachyonic strings.  It corresponds to
the $ISO(2,1)$ orbit with $ <{\cal K},{\cal K}>=0$         and
$({\cal K},{\cal K})=-2n_in^i\;.$  Here we get that
  \begin{eqnarray}
{}^g[{\cal K}_1]  &=& \epsilon^{ijk} (\Lambda n)_k \; t_i \otimes t_j  -
 \epsilon^{ijk} (\Lambda n)_\ell x^\ell x_k \; u_i \otimes u_j\cr
& &+\;(\Lambda n)_\ell x^\ell \;(t_i \otimes u^i  - u_i \otimes t^i)
+(\Lambda n)^i x^j \;(u_i \otimes t_j  - t_j \otimes u_i) \;.
\end{eqnarray}
Using this and the expression for the right invariant Maurer-Cartan
form
\Eq{dg g^{-1}={1\over2}\epsilon^{ijk}(d\Lambda\Lambda^{-1})_{ij} t_k +
[dx^i -(d\Lambda\Lambda^{-1}x)^i]u_i\;,}
we compute the following currents:
 \begin{eqnarray}
P_i^\alpha=P_{(1)i}^\alpha &=&\epsilon^{\alpha\beta} \biggl(
{1\over 2}\epsilon_{mjk}(\partial_\beta\Lambda\Lambda^{-1})^{jk}
 [\delta^m_i  (\Lambda n)_\ell x^\ell     -(\Lambda n)^m x_i ] \cr
& & \qquad \;+\;\epsilon_{ijk}[\partial_\beta x^j-(\partial_\beta \Lambda
   \Lambda^{-1} x)^j] (\Lambda n)^k  \biggr)\;,  \\   & &\cr
J_i^\alpha=J_{(1)i}^\alpha& =&-\epsilon^{\alpha\beta} \biggl(
(\partial_\beta\Lambda\Lambda^{-1}{)_i}^j(\Lambda n)_\ell x^\ell x_j \cr
& &\qquad \;+\; [ \partial_\beta x^j-(\partial_\beta \Lambda
   \Lambda^{-1} x)_j ]
[\delta^j_i  (\Lambda n)_\ell x^\ell    -(\Lambda n)_i x^j ]  \biggr)\;.
\end{eqnarray}
These two currents can be shown to be related by
\Eq{J_{(1)i}^\alpha-\epsilon_{ijk} x^j P_{(1)}^{k\alpha}=0\;.\label{oam}}
We therefore argue that for such strings the angular momentum current
consists only of an orbital term, and that no spin is present.  This
is a well known result for the Nambu string.

\underbar{Case 2}:  ${\cal K}=
{\cal K}_{2}={1\over 2}\chi^{ij} \;( u_i \otimes t_j-
t_j \otimes u_i)$.  Since the action (\ref{stac})
 is antisymmetric with respect to
the exchange of the two vector spaces, this is the most general ansatz
for the tensor ${\cal K}$ which is linear in $u_i$ and in $t_j$.
It corresponds to
the $ISO(2,1)$ orbit with $ <{\cal K},{\cal K}>=\chi_{ij}\chi^{ij} $
 and       $({\cal K},{\cal K})=0\;.$  From it we get
\begin{eqnarray}
 {}^g[{\cal K}_{2}] & =& {1\over 2}(\Lambda\chi\Lambda^{-1})^{ij}
 \;( u_i \otimes t_j-t_j \otimes u_i)  \cr
& & + \; {1\over 2} [{\rm tr} \chi\; x_k -
(\Lambda\chi\Lambda^{-1}{)^\ell}_k
x_\ell]\;\epsilon^{ijk}\; u_i \otimes u_j \;.\end{eqnarray}
We now obtain the following currents:
\begin{eqnarray}
P_i^\alpha&=&P_{(2)i}^\alpha=-{1\over 4}
\epsilon^{\alpha\beta} \epsilon_{jk\ell}
(\Lambda^{-1}\partial_\beta\Lambda)^{jk} (\chi \Lambda^{-1} {)^\ell}_i
 \;,\\   & & \cr
J_i^\alpha&=&J_{(2)i}^\alpha={1\over 2}\epsilon^{\alpha\beta} \biggl(
[\Lambda \chi \partial_\beta(\Lambda^{-1} x)]_i - [
{\rm tr} \chi \; (\partial_\beta\Lambda\Lambda^{-1})_{ji} +
(\Lambda \chi \partial_\beta\Lambda^{-1})_{ji}] x^j  \biggr)\;.
\end{eqnarray}
Now the analogue of eq. (\ref{oam}) is no longer true, i.e.
\Eq{S_{i}^\alpha = J_{(2)i}^\alpha-\epsilon_{ijk} x^j P_{(2)}^{k\alpha}
\ne 0\;.\label{spn}}
We then conclude that a spin current is present in this case.

\underbar{Case 3}:  ${\cal K}=
{\cal K}_{3} =\epsilon^{ijk} \;\nu_k\; u_i \otimes u_j$.
Here both invariants vanish,
$ <{\cal K},{\cal K}>=0 $
 and $ ({\cal K},{\cal K})=0\;.$
The currents $P_i^\alpha$ and $J_i^\alpha$  are
trivially conserved in this case.  This is because
the integrand in (\ref{stac})  can be expressed as an exact two form
on $ISO(2,1)$:
\Eq{<{\cal K}, g^{-1}dg \otimes g^{-1}dg >   = {1\over 2} \epsilon_{ijk}
(\Lambda^{-1} d\Lambda)^{ij}(\Lambda^{-1} d\Lambda\;\nu)^k =
    -\; d\;\epsilon_{ijk}\;\nu^k (\Lambda^{-1} d\Lambda)^{ij} \;.   }
Although this term does not contribute to the classical equations
 of motion, it can affect the quantum dynamics.  Furthermore, it
is known to be associated with the $\theta$-vacua of
string theory.\cite{bal},\cite{theta}

\underbar{Case 4}:  ${\cal K}= {\cal K}_1 + {\cal K}_{2} $.  This defines
the most general classical system with
the string action given by (\ref{stac}).  It therefore contains
the case of the Nambu string.  Now both invariants can be nonzero,
 $<{\cal K},{\cal K}>=\chi_{ij}\chi^{ij}$ and
$ ({\cal K},{\cal K})=-2n_in^i\;.$
The conserved currents are now  given by:
\begin{eqnarray}
P_i^\alpha&=&P_{(1)i}^\alpha +P_{(2)i}^\alpha \;,  \\
J_i^\alpha&=&J_{(1)i}^\alpha +J_{(2)i}^\alpha \;.
\end{eqnarray}
For such strings, we can identify both an orbital and a spin
 angular momentum current, i.e.
$J_i^\alpha=L_{i}^\alpha +S_{i}^\alpha \;. $  The spin current
$S_{i}^\alpha$ is defined  in eq. (\ref{spn}), while the orbital angular
momentum   $L_{i}^\alpha$   is given by
\Eq{L_{i}^\alpha = J_{(1)i}^\alpha
+\epsilon_{ijk} x^j P_{(2)}^{k\alpha}\;.}

From the above discussion we conclude that a spin current
is present for the case of $ISO(2,1)$ orbits with $<{\cal K},{\cal K}>
\ne 0$.   If we include a Wess-Zumino term as is done in ref. \cite{bal}
an additional term of the form $\epsilon_{ijk}\epsilon^{\alpha\beta}
(\Lambda^{-1}\partial_\beta\Lambda)^{jk}$ contributes to the angular
momentum current.

It is easy to  embed our spinning strings in curved space-time.
Now the action should be invariant under local Poincar\'e transformations,
\Eq{g\rightarrow h_L \circ g \;,\label{lpt}}
 where like $g$, $h_L$ are functions
on the two dimensional world sheet, taking values in  $ISO(2,1)$.
We recall that the action (\ref{stac}) was instead
invariant under global Poincar\'e transformations.
To elevate it to a local invariance, we replace $g^{-1}dg$ by
\Eq{ g^{-1}Dg = g^{-1}dg  +{}^{g^{-1}}[ A] \;,\label{cmcf}}  where
 $A=\omega^it_i +e^iu_i$ are the connection one-forms for $ISO(2,1)$
evaluated on the string world sheet.  Under
 Poincar\'e gauge transformations (\ref{lpt}),
\Eq{ A\rightarrow {}^{h_L}[A] - dh_L {h_L}^{-1}\;,} and as a result
$ Dgg^{-1} $  is invariant.  Then
\Eq{{\cal S}_{string} =  \int <{\cal K}, g^{-1}Dg \otimes g^{-1}Dg >\;.
\label{staig} } is gauge invariant, and hence gives the string action
in curved space-time.

The equations of motion  obtained by varying $g$ now state that
the momentum  and angular momentum currents
are covariantly conserved.  To see this we can again use (\ref{dmc})
along with $\delta({}^{g^{-1}}[A]) = {}^{g^{-1}}[A,\epsilon]\;.$  Then
\Eq{\delta {\cal S}_{string}
=2\int <{\cal K}, {}^{g^{-1}}[D\epsilon]\otimes
g^{-1}Dg> =2\int <{}^g[{\cal K}], D\epsilon\otimes Dgg^{-1}>\;,}
where we have used the invariance property
of the scalar product, $D\epsilon = d\epsilon +[A,\epsilon]$ and
the identity  $     {}^{g^{-1}}[Dg g^{-1}]= g^{-1} Dg \;.$
Upon integrating by parts we now arrive at the equations of motion
\Eq{ d\; <{}^g[{\cal K}], T_A\otimes Dgg^{-1}>-
<{}^g[{\cal K}], [A\otimes \BI,T_A\otimes Dgg^{-1}]>
=0\;,}  where the $T_A$'s once again
denote the  generators of $ ISO(2,1) $.  We then get the following
 generalizations of (\ref{cop}) and (\ref{coj}):
\begin{eqnarray}
 \partial_\alpha {\cal P}_i^\alpha+\epsilon_{ijk}\omega_\alpha^j
 {\cal P}^{\alpha k}&=&0\;,\qquad\;
{\cal P}_i^\alpha=\epsilon^{\alpha\beta} <{}^g[{\cal K}], u_i\otimes
{\cal D}_\beta gg^{-1}>\;,\\
  \partial_\alpha {\cal J}_i^\alpha+\epsilon_{ijk}(\omega_\alpha^j
 {\cal J}^{\alpha k}+   e_\alpha^j  {\cal P}^{\alpha k} )&=&0\;,\qquad
{\cal J}_i^\alpha=\epsilon^{\alpha\beta} <{}^g[{\cal K}], t_i\otimes
{\cal D}_\beta gg^{-1}>\;,\end{eqnarray}  $ \omega_\alpha^j $,
$e_\alpha^j $ and   ${\cal D}_\beta gg^{-1}$ denoting the
world sheet components of the one forms $ \omega^j $,
$e^j $ and     $ Dgg^{-1} $, respectively.    The currents play the
role of sources for the $ISO(2,1)$ curvature.  For this we can take
the Chern-Simons action\cite{wit} for the fields.  Then
${\cal P}_i^\alpha$ is a source for the $SO(2,1)$ curvature, while
${\cal J}_i^\alpha$ is a source for the torsion.

\bigskip
\bigskip

{\parindent 0cm{\bf Acknowledgements:}}
We are grateful for discussions with R. Casadio.
We were supported in part
by the Department of Energy, USA, under contract number
DE-FG05-84ER40141.

\end{document}